\documentclass{emulateapj}
\newcommand{\msun}{$M_{\odot}$}

\newcommand{\wds}{white dwarf}

\newcommand{\gmos}{{\em GMOS}}
\newcommand{\wfpc}{{\em WFPC2}}
\newcommand{\lris}{{\em LRIS}}

\newcommand{\gc}{globular cluster}

\shorttitle{White dwarf spectra in M4}
\shortauthors{Davis et al.}
\begin{document}
%%fakesection  title, authors, and affiliations
\title{The spectral types of white dwarfs in Messier 4}
\author{
D.~Saul Davis\altaffilmark{1},
Harvey B.~Richer\altaffilmark{1},
R.~Michael Rich\altaffilmark{2}, 
David R.~Reitzel\altaffilmark{2},
Jason S.~Kalirai\altaffilmark{3}
}
\altaffiltext{1}{University of British Columbia, Vancouver, Canada;
sdavis@astro.ubc.ca}
\altaffiltext{2}{Division of Astronomy University of California at Los Angeles,
Los Angeles, CA. 90095}
\altaffiltext{3}{Space Telescope Science Institute, Baltimore, MD 21218}

%%fakesection  abstract and keywords
\begin{abstract}
We present the spectra of 24 white dwarfs in the direction of the globular
cluster Messier 4 obtained with the {\em Keck/LRIS} and {\em Gemini/GMOS}
spectrographs.  Determining the spectral types of the stars in this sample, we
find 24 type DA and 0 type DB (i.e., atmospheres dominated by hydrogen and
helium respectively).  Assuming the ratio of DA/DB observed in the field with
effective temperature between $15\,000$--$25\,000$ K, i.e., 4.2:1, holds for
the cluster environment, the chance of finding no DBs in our sample due simply
to statistical fluctuations is only $6\times10^{-3}$. The spectral types of the
$\sim100$ white dwarfs previously identified in open clusters indicate that
DB formation is strongly suppressed in that environment. Furthermore, all the
$\sim10$ white dwarfs previously identified in other globular clusters are
exclusively type DA. In the context of these two facts, this finding suggests
that DB formation is suppressed in the cluster environment in general. Though
no satisfactory explanation for this phenomenon exists, we discuss several
possibilities.
\end{abstract}
\keywords{(Galaxy:) globular clusters: individual (Messier 4),
(stars:) white dwarfs.}

\section{Introduction}
Because of their intrinsically faint luminosities, white dwarfs are relatively
hard to study spectroscopically. It is now generally accepted that the number
of DA to non-DA white dwarfs is a function of temperature \citep[see][for
example]{brl97}. Because a given \wds\ will be born with a high temperature,
and subsequently cool through the entire range of observed white-dwarf
temperatures, the implication of this observation is that white dwarfs can
change their spectral type multiple times throughout their evolution. This
means that, unlike main-sequence stars, \wds s that are physically similar can
have very different spectra, or vice-versa. The dominant physical variable that
controls the spectral type of a \wds\ during its evolution is the mass of the
very-thin atmosphere layer.

\citet{hl03} outline the presumed multiple evolutionary sequences of white
dwarfs with ``thick'' and ``thin'' hydrogen envelopes, with envelope masses of
$10^{-4}\,M_\odot$ and $\sim10^{-10}$\msun\ respectively.  Stars with thick
hydrogen layers have atmospheres rich in hydrogen throughout their evolution,
and will always appear as DAs (exhibiting H {\sc i} lines). White dwarfs with
thin hydrogen layers have a much more complicated evolution, and may at varying
times have atmospheres dominated by either hydrogen or helium.  These \wds s
may be born with DA or DO (exhibiting He {\sc ii} lines) spectral types.  As
the stars cool, helium ions will recombine, and the DOs will transform into
DBs (exhibiting He {\sc i} lines). 

At temperatures between $30\,000$--$150\,000$ K, DAs far outnumber DBs. In
fact, from $30\,000$--$45\,000$ K there is the so-called ``DB gap''
\cite{lwh86}.  This temperature range was, until recently, completely devoid of
DB stars.  With DR4, the SDSS had accumulated $10^4$ white-dwarf spectra, and
within this sample, approximately 10 DB white dwarfs were found in this
temperature range \citep{elk06}.  While this temperature range is no longer
strictly a ``gap'', it is still true that the DA/DB ratio is approximately 2.5
times greater at $30\,000$ K than it is at $20\,000$ K.  This implies that an
atmospheric transformation takes place in approximately 10\% of DAs as they
cool through this range \citep{elk06}.

Below $30\,000$ K, a helium convection zone is established. Convective
velocities can become high enough to overshoot, and mix the helium layer with
the hydrogen atmosphere, converting a type DA to a type DB.  This occurs in
$\sim25$\% of white dwarfs in this temperature range in the field \citep{hl03}.

For temperatures lower than $6\,000$ K, the situation becomes even more
complicated and uncertain \citep{brl97}.  At these low temperatures the variety
of spectral types increases, but in the non-DA gap ($6\,000$--$ 5\,000$ K)
non-DA white dwarfs have yet to be observed.  As \wds s cool below $5\,000$ K,
neither hydrogen nor helium lines are excited, and the resultant spectrum is
almost featureless, leading to the DC spectral classification. Even more rare
are white dwarfs showing metal lines. These are the DQs, for those showing
carbon features, and the DZs, for other atomic species.  Finally, DP and DH
white dwarfs show evidence of having polarized and non-polarized magnetic
fields respectively.

Prior to the year 2000, the spectra of cluster white dwarfs had been obtained
in a piecemeal fashion---only one or two stars would be studied in each
cluster, typically by different authors. The Canada-France-Hawaii Telescope
(CFHT) Open Star Cluster Survey changed this. The idea behind this survey was
to obtain high-quality, wide-field images of many open clusters.  The rich
clusters (i.e., those with well-populated white dwarf cooling sequences) could
then be identified, and followed up spectroscopically.  Previously,  when
spectral types were obtained for only several white dwarfs at a time, the
absence of a particular spectral type was neither particularly surprising nor
interesting.  

\citet{krh05} reported 21 spectral identifications in NGC 2099.  Assuming the
same DA/DB ratio observed in the field for this temperature range ($15\,000$
K--$30\,000$ K) held for this cluster, it was expected that several DB white
dwarfs would be found.  Contrary to expectations, no DB white dwarfs were
found.  According to \citet{krh05}, finding a DB/DA ratio of 0/21 would occur
approximately 2\% of the time simply due to statistical fluctuations.  This
finding prompted \citet{krh05} to examine all white dwarf spectral
identifications in young open clusters. These cluster were all very young. In
fact, NGC 2099 was the oldest of the sample. Because turn-off mass (and
therefore white-dwarf mass) is inversely correlated with cluster age, all of
the white dwarfs in this sample are very massive. Of all the 65 white dwarfs
that had been spectroscopically identified in young open clusters at that
point, all were of type DA.  This had a vanishingly small chance of occurring
due to a statistical fluctuation under the hypothesis that the same DA/DB ratio
held as in the field.  

An obvious explanation for this observation was not apparent.  One early
explanation was the re-accretion of residual intra-cluster gas in open
clusters.  However, this explanation appeared not to hold up on closer
examination. The escape velocity in open clusters is typically very low
($v_{\rm esc} \sim1\, km\, s^{-1}$), and gas ejected by stellar winds ($v_{\rm
wind} \sim 10\, km\, s^{-1}$) would be expected to quickly escape. It was found
that the accretion rate of intra-cluster gas in open clusters should be no
greater than that of the ISM in the disk. \citet{krh05} postulated an
explanation based on the mass of the white dwarfs.  Because NGC 2099 is a
young cluster, the young white dwarfs are high mass ($\sim0.8$ \msun)
compared to the average disk population ($\sim0.6$ \msun).  He convection is
inhibited in high-mass white dwarfs, and therefore even white dwarfs with thin
hydrogen atmospheres would remain type DA in this temperature range.

Globular clusters are the oldest Galactic stellar population, and should
therefore create the lowest-mass white dwarfs ($\sim0.5$ \msun).  It is of
interest to determine whether the same paucity of DB white dwarfs exists in
globular clusters. Recent observations suggest that the same paucity is
indeed there. In 2004, \citet{mkz04} obtained spectra of $5$ white dwarfs in
NGC 6752 and another $4$ in NGC 6397, all of which were found to be type DA.
Assuming that the DA/DB ratio is the same as in the field, the chance of
finding no DB white dwarfs given the number observed in NGC 6752 and NGC 6397
is 0.35 and 0.43 respectively. 

The current paper follows from an ultra-deep {\em HST/WFPC2} study of the
nearest globular cluster, Messier 4 (NGC 6121) by \citet{rbf02}. White dwarfs
identified in this photometry were followed up with the {\em Gemini/GMOS} and
{\em Keck/LRIS} spectrographs. The primary science goal of the spectrographic
follow-up was the determination of the masses of the white dwarfs (see Kalirai
et al. 2009). While spectroscopic masses were determined for only a subset of
the observed \wds s due to signal-to-noise ratio constraints, the spectral
types were determined for 24 of the 25 candidates with a high-probability of
being a white dwarf. The determination of the spectral type of 24 \wds s more
than doubles the number of existing identifications of white dwarfs in globular
clusters. As in the earlier studies, we find a complete lack of DBs in our
sample.

\section{Photometric target selection} 
A critical, and in this case particularly challenging,  component of any
spectroscopic study, is the target selection. We used two sources of
pre-imaging to select stars that lie on the white-dwarf cooling curve of
Messier 4. White dwarf candidates were selected from the \wfpc\ photometry
published in \citet{rfb04}, and from {\em Gemini/GMOS} imaging.  These targets
were followed up with the two multi-slit spectrographs: {\em Gemini/GMOS} and
{\em Keck/LRIS}. 

Our most secure white dwarf targets come from the \wfpc\ photometry. Stars in
these fields have their cluster membership confirmed by their proper motions.
However, there are two limitations with the \wfpc\ imaging. First, both the
\gmos\ and the \lris\ fields-of-view are much larger than the \wfpc\ field. In
order to maximize the number of spectra, we must distribute the targets evenly
over the spectroscopic fields, and therefore some of our spectroscopic targets
fell outside the \wfpc\  field of view.  Second, the PSFs of ground-based
instruments are obviously much broader than that of the \wfpc. Some of the
white dwarfs that are easily resolved with \wfpc\  are lost in the scattered
light from bright stars in ground-based photometry in this crowded field.

\subsection{\gmos\ target selection}
\begin{figure}
\includegraphics[width=\columnwidth]{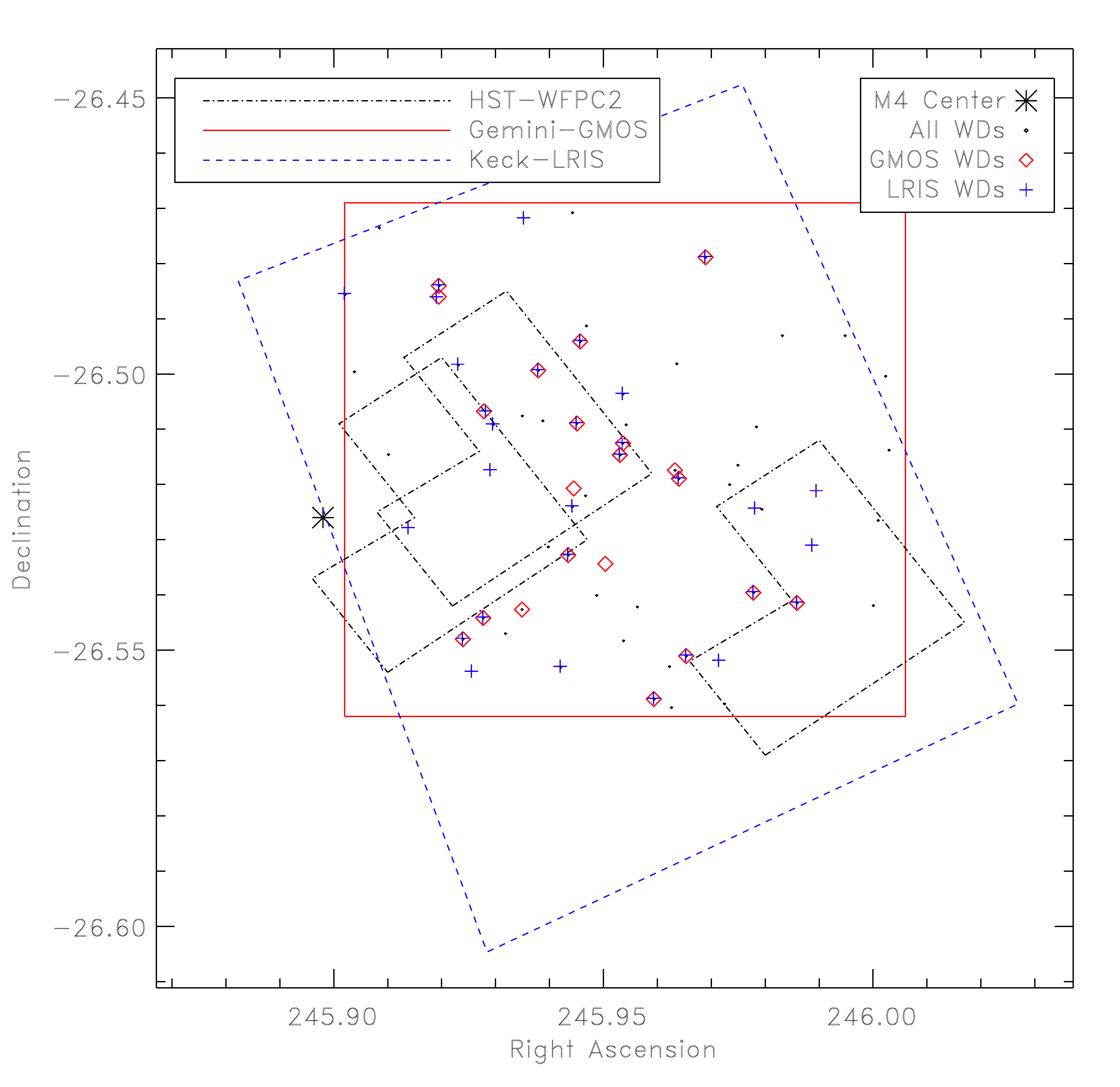} \caption{The
footprint of the HST/WFPC2, Gemini/GMOS, and Keck/LRIS instruments.  The white
dwarf candidates selected from the \gmos\ photometry (i.e., those stars with 21
$<$ F555W $<$ 25, and F555W-F814W $<$ 0.9.) are shown as dots.  The objects
targeted for \gmos\ spectroscopy are shown with diamonds. The objects targeted
for \lris\ spectroscopy are shown with pluses. \label{foot.fig}}
\end{figure}

In order to obtain stars over the entire \gmos\ field, \gmos\ pre-imaging was
obtained. The stars selected from the \gmos\ photometry could only be selected
in colour-magnitude space, and not proper-motion space, and therefore their
cluster membership is less certain. \lris, \gmos, and \wfpc\ all have different
fields of view, and some areas of the sky have been studied with all three
instruments, while others have only been studied by one or two.  The footprints
for the various instruments are shown in Figure \ref{foot.fig}.

In order to effectively select white dwarf candidates over the entire \gmos\
field, pre-imaging was taken in both the g$^\prime$ and r$^\prime$ filters.
The images were corrected for bias and flat fielding by the {\em Gemini}
pipeline. The pre-processed images were then reduced with the standard
DAOPHOT/Allstar reduction techniques \citep{ste87}. This photometry was not
used for anything other than the target selection, and hence was not rigorously
calibrated.  There are $10^3$ stars imaged both with \wfpc\ and \gmos.  In
order to calibrate the \gmos\ photometry, a transformation between the \gmos\
and the calibrated \wfpc\ photometry was determined. The transformation was of
the form:
\begin{eqnarray}
	g - r & = & g^\prime - r^\prime - x_0  \nonumber \\
	    g & = & g^\prime - y_0  \nonumber \\
\text{F555W}  & = & y_0 + y_1\, g + y_2\, (g-r) + y_3\, (g-r) g \nonumber  \\
\text{F555W-F814W} & = & x_0 + x_1\, g + x_2\, (g-r) + x_3\, (g-r) g  \nonumber 
	\label{gtogp}
\end{eqnarray}
The coefficients were determined from minimizing $\chi^2$ using the
downhill-simplex method {\em Amoeba} \citep{pft86}, and were found to be:
$x_0=19.0$, $x_1=1.72\times10^{-3}$, $x_2=1.19\times10^{-1}$,
$x_3=1.12\times10^{-2}$, $y_0=19$, $y_1=-2.02\times10^{-1}$,
$y_2=-4.59\times10^{-3}$, and $y_3=-5.67\times10^{-2}$.

\begin{figure*}
\includegraphics[width=\textwidth]{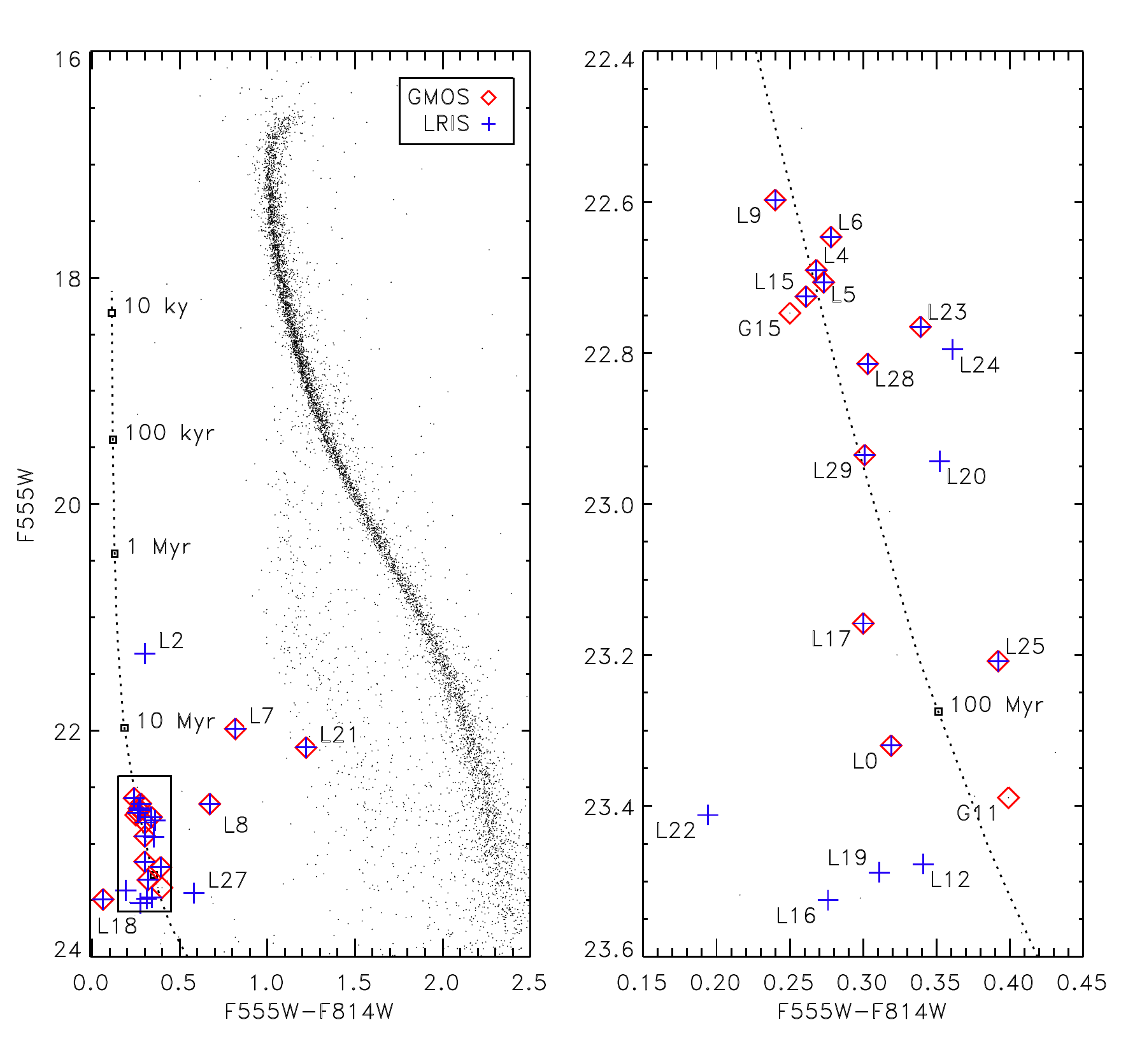}
\caption{The CMD constructed from the \gmos\ photometry. There is a clear white
dwarf cooling sequence extending from $22.5<$F555W$<24$ with an approximate
colour of F555W-F814W=0.3.  The left panel shows the entire range, while the
right panel shows only the white dwarf cooling sequence. A cooling sequence
(assuming (m-M)$_{\rm F814}$=12.49 and E(F555W-F814W)=0.06) of a 0.5 \msun\ DA
white dwarf is shown \citep{fbb01}. The \gmos\ photometry in the \wds-region
has been replaced by \wfpc\ photometry where it exists.
\label{cmd.fig}}
\end{figure*}

The \gmos\ photometry was then transformed to the \wfpc\ bands. The CMD
constructed from the \gmos\ photometry is presented in Figure \ref{cmd.fig}. A
clear white dwarf cooling sequence can be seen beginning at F555W=22.5 and
F555W-F814W=0.30.

Figure \ref{foot.fig} shows the location of the white dwarf candidates selected
from the \gmos\ photometry (points). Note the paucity of \gmos\ points in the
inner \wfpc\ fields. Though there are many white dwarfs here, the crowding is
such that they become very difficult to detect with ground-based telescopes,
such as Gemini. The \gmos\ points are those stars with magnitudes between
F555W=22.5 and F555W=24.5 and colours less than F555W-F814W=0.9. All the
relevant candidate selection measurements are tabulated in Table
\ref{targtab.tab}.

\begin{deluxetable*}{ccccccccc}
\tablecaption{\label{targtab.tab}The measurements used for target selection and measured spectral type}

\tablecolumns{8}
\tablehead{
           \colhead{LRIS \#} &
           \colhead{GMOS \#} & 
           \colhead{RA} &
           \colhead{DEC} &
           \colhead{F555W-F814W} &
           \colhead{F555W}       &
           \colhead{F555W-F814W} &
	   \colhead{F555W}      \\
	   \colhead{}        &
	   \colhead{}        &
	   \colhead{(J2000)}        &
	   \colhead{(J2000)}        &
	   \multicolumn{2}{c}{(\gmos)} &
	   \multicolumn{2}{c}{(\wfpc)} }
\startdata
LRIS-0  & GMOS-21 & 245.9579 & -26.5589 & 0.319 & 23.320 &   --- &    ---  \\ % 2314
LRIS-2  &     --- & 245.9406 & -26.5531 & 0.302 & 21.321 &   --- &    ---  \\ % 3493
LRIS-4  & GMOS-20 & 245.9638 & -26.5511 & 0.268 & 22.690 &   --- &    ---  \\ % 1956
LRIS-5  & GMOS-19 & 245.9224 & -26.5480 & 0.273 & 22.706 &   --- &    ---  \\ % 5072
LRIS-6  & GMOS-18 & 245.9262 & -26.5442 & 0.278 & 22.646 &   --- &    ---  \\ % 4731
    --- & GMOS-15 & 245.9334 & -26.5426 & 0.250 & 22.747 &   --- &    ---  \\ % 4094
LRIS-7  & GMOS-16 & 245.9844 & -26.5414 & 0.818 & 21.985 &   --- &    ---  \\ %  838
LRIS-8  & GMOS-13 & 245.9763 & -26.5396 & 0.671 & 22.649 &   --- &    ---  \\ % 1250
LRIS-9  & GMOS-14 & 245.9419 & -26.5328 & 0.258 & 22.502 & 0.240 & 22.597  \\ % 3390
LRIS-12 &     --- & 245.9427 & -26.5241 & 0.216 & 23.518 & 0.341 & 23.477  \\ % 3331
    --- & GMOS-11 & 245.9617 & -26.5174 & 0.399 & 23.389 &   --- &    ---  \\ % 2069
LRIS-15 & GMOS-10 & 245.9625 & -26.5190 & 0.261 & 22.725 &   --- &    ---  \\ % 2033
LRIS-16 &     --- & 245.9276 & -26.5175 &  ---  &  ---   & 0.276 & 23.525  \\ %
LRIS-17 & GMOS-6  & 245.9515 & -26.5147 & 0.323 & 23.200 & 0.300 & 23.158  \\ % 2715
LRIS-18 & GMOS-8  & 245.9521 & -26.5125 & 0.097 & 23.891 & 0.063 & 23.492  \\ % 2680
LRIS-19 &     --- & 245.9281 & -26.5092 & 0.431 & 23.329 & 0.311 & 23.488  \\ % 4539
LRIS-20 & GMOS-7  & 245.9436 & -26.5090 & 0.318 & 23.010 & 0.352 & 22.943  \\ % 3271
LRIS-21 & GMOS-9  & 245.9263 & -26.5067 & 1.221 & 22.150 &   --- &    ---  \\ % 4712
LRIS-22 &     --- & 245.9521 & -26.5037 & 0.194 & 23.412 &   --- &    ---  \\ % 2675
LRIS-23 & GMOS-4  & 245.9364 & -26.4994 & 0.513 & 22.536 & 0.339 & 22.765  \\ % 3826
LRIS-24 &     --- & 245.9216 & -26.4984 & 0.262 & 22.724 & 0.361 & 22.795  \\ % 5153
LRIS-25 & GMOS-3  & 245.9442 & -26.4941 & 0.392 & 23.208 &   --- &    ---  \\ % 3217
LRIS-27 &     --- & 245.9006 & -26.4856 & 0.582 & 23.437 &   --- &    ---  \\ % 7326
LRIS-28 & GMOS-2  & 245.9179 & -26.4840 & 0.303 & 22.814 &   --- &    ---  \\ % 5484
LRIS-29 & GMOS-1  & 245.9674 & -26.4789 & 0.301 & 22.935 &   --- &    ---     % 1748
\enddata
\end{deluxetable*}

\subsection{Keck target selection}
We constructed a CMD from \lris\ photometry, but it lacks a coherent
white-dwarf-cooling sequence. Furthermore, by the time we were selecting
targets for the \lris\ mask, the \gmos\ spectroscopy had already been
performed. We therefore had spectroscopic confirmation of many white dwarfs.
The selection for \lris\ was therefore performed with the same inputs as that
for the \gmos\ photometry. The locations of the \lris\ targets shown in the
\gmos\ colour-magnitude space are shown in Figure \ref{cmd.fig}, and their
astrometry in Figure \ref{foot.fig}.

\section{Spectral Reductions}
\subsection{GMOS Reductions}
In total, we obtained approximately 14.5 and 9 hours of science exposure with
\gmos\ in 2005A and 2006B respectively. Because of the queue system we were
able to require all exposures to be obtained in sub-arcsecond seeing. The
\gmos\ observations obtained are listed in Table \ref{obstime.tab}.  Our \gmos
mask consisted of 21 objects. The slits were all 0\farcs8 wide and
5\farcs0 long. We used the B1200 grating, and binned by two pixels in the
spectral direction resulting in a resolution of $R=1900$.

\begin{deluxetable*}{ccccll|cccccl}
	\tablecaption{\label{obstime.tab} Observing information}
\tablecolumns{12}
\tablehead{
	   \multicolumn{5}{c}{\gmos} &
	   \colhead{}            &
	   \colhead{}            &
	   \multicolumn{5}{c}{\lris} \\
           \colhead{exposure}    &
           \colhead{date}        &
           \colhead{number of}   &
           \colhead{seeing}      &
           \colhead{airmass}     &
	   \colhead{}            &
	   \colhead{}            &
           \colhead{exposure}    &
           \colhead{date}        &
           \colhead{number of}   &
           \colhead{seeing}      &
           \colhead{airmass}     \\
           \colhead{time (s)}    &
           \colhead{}            &
           \colhead{exposures}   &
	   \colhead{(\arcsec)}   &
	   \colhead{}            &
	   \colhead{}            &
	   \colhead{}            &
           \colhead{time (s)}    &
           \colhead{}            &
           \colhead{exposures}   &
	   \colhead{(\arcsec)}   &
	   \colhead{}            }
\startdata
3600 & 06/06/2005 & 2 & 0.8 & 1.08--1.23 & & & 2700 & 04/21/2007 & 2 & 0.9 & 1.45--1.49 \\
3600 & 06/07/2005 & 3 & 0.8 & 1.04--1.34 & & & 900  & 04/21/2007 & 1 & 1.0 & 1.64 \\
1687 & 06/08/2005 & 1 & 0.8 & 1.45       & & & 2700 & 04/22/2007 & 3 & 1.0 & 1.45--1.52 \\
3600 & 06/09/2005 & 5 & 0.8 & 1.02--1.45 & & & 1800 & 04/22/2007 & 1 & 1.0 & 1.56 \\
3600 & 08/04/2005 & 2 & 0.8 & 1.13--1.33 & & & 1800 & 07/14/2007 & 2 & 1.0 & 1.45--1.63 \\
3600 & 08/08/2005 & 1 & 0.8 & 1.03       & & & 2100 & 07/14/2007 & 1 & 1.2 & 1.52 \\
620  & 08/08/2005 & 1 & 0.8 & 1.14       & & & 1800 & 07/15/2007 & 4 & 0.8 & 1.45--1.51 \\
3600 & 08/09/2005 & 1 & 0.8 & 1.03--1.07 & & & 1800 & 04/12/2008 & 5 & 2.0 & 1.45--1.52 \\
3600 & 08/19/2006 & 3 & 0.8 & 1.03--1.33 & & & \\
3600 & 08/20/2006 & 1 & 0.8 & 1.12       & & \\
3600 & 08/21/2006 & 2 & 0.8 & 1.06--1.18 & & \\
3600 & 09/15/2006 & 1 & 0.8 & 1.23 & & \\
3600 & 09/16/2006 & 1 & 0.8 & 1.25 & & \\
3600 & 09/20/2006 & 1 & 0.8 & 1.32 & & 
\enddata
\end{deluxetable*}

The raw data frames were downloaded from the Canadian Astronomy Data Center
(CADC) in multi-extension FITS (MEF) format.  We reduced the data using the
Gemini {\sc iraf} Package, version 1.4. \gmos\ is composed of three separate
chips.  The dispersion axis is perpendicular to the long axis of the chips, and
therefore the dispersed spectra will cross the gaps between the chips, leaving
gaps in the spectral coverage. The precise spectral coverage for a given star
depends on its position in the detector, and is therefore slightly different
for each star.  Because the spectral range is different for each star, we are
unable to choose a central wavelength such that no star will have a Balmer line
that falls on a chip gap. To avoid a Balmer line falling on a gap and rendering
the spectrum useless, we obtain the spectrum at two different grating offsets.
This is equivalent to dithering a camera when obtaining imaging. The central
wavelength was shifted from the default value of 4620~${\rm \AA}$ to 4720~${\rm
\AA}$ for half of the exposures.  These two sets of spectra are handled
separately until the wavelength calibration is determined.

The \gmos\ data were reduced using the {\sc iraf/Gemini} package. For the
initial stages of the reductions, the standard steps, i.e., {\sc gprepare}
(updates FITS header information, and associates image with mask definition
file), {\sc gbias} (applies the bias correction), {\sc gflatten} (corrects for
the pixel-to-pixel gain variations), and {\sc gsreduce} (subtracting the
overscan, cleaning the image for cosmic rays, mosaicing the three chips
together, interpolating the pixels with within the chip gap, and cutting the
MOS slits into separate spectra).  For the wavelength calibration, we obtained
multiple spectroscopic frames from a CuAr lamp exposure.  The automatic
wavelength fitting routine, {\sc gswavelength}, was good enough to find a rough
wavelength calibration, however, the calibration had to be verified
interactively.  The resultant residuals in the fit to the emission lines were
all well behaved, typically at the 0.4 ${\rm \AA}$ level.  With this template,
we then used the {\sc gstransform} task to apply the wavelength calibration to
the science frames. 

The sky subtraction was completed with the {\sc gsskysub} task.  For the \gmos\
data, the star was always centered on the slit. The central $\sim1$\farcs0 of
each slit was assumed to contain the stellar signal, and the sky was estimated
from the remaining pixels.  This typically gave us 6 pixels for the star, and
11 pixels both above and below the star for the sky estimation (the stars are
centered in the slits).  There are only 37 spatial pixels due to our factor of
two binning in this direction. A number of pixels at both the top and bottom of
each slit were discarded.  

Due to the faintness of our stars, the automatic routines for extracting the
spectra were generally unsuccessful.  We manually extracted each of the spectra
into a 1D format using the {\sc gsextract} task.  This was a particularly
challenging and uncertain aspect of the reductions.  The sensitivity of \gmos\
decreases to wavelengths bluer than H$\varepsilon$, and the already faint
stars became almost invisible.  Delineating the trace in these parts of the
spectra was very uncertain.  Unfortunately, in order to determine a reliable
spectroscopic mass (the primary science goal of this observing project),
H$\varepsilon$ is crucial, and H$8$ is very useful. The lack of flux at these
blue wavelengths ultimately limited the utility of the \gmos\ spectra for
constraining the mass of the objects spectroscopically,  but were more than
sufficient to determine a spectral type. The resulting manually extracted
spectra were combined in a weighted average based on the signal-to-noise of
each spectrum.

\subsection{LRIS Reductions} 
We also obtained multi-object spectroscopy using \lris\ on the Keck I
telescope. The \lris mask was designed with 29 slits. Two masks, with differing
slit width (0\farcs8 and 1\farcs), were cut in order to match the seeing on a
given night. The slits were not uniform in length, but the minimum slit length
was 5\farcs0. We used the 400/3400 grism resulting in a resolution of $R=1000$.

The target selection was performed using \gmos\ photometry, and used
the information from the reduced \gmos\ spectra mentioned in the previous
section.  The reduction of the \lris\ data was performed entirely within {\sc
iraf}.  In total, we were granted 7 half-nights with \lris, however, due to
poor weather conditions, only 10.6 hours of usable exposure time were obtained.
The observations obtained are listed in Table \ref{obstime.tab}. The data were
of highly variable quality.  The observations were collected with seeing
ranging from 0\farcs8 to over 2\farcs0. In an uncrowded field, the
signal-to-noise ratio for equal exposure times will be higher for observation
obtained in good seeing.  In a crowded field this effect is exacerbated due to
the scattered light from nearby bright stars that is incident upon the slit in
poor seeing conditions. The final signal-to-noise ratios of the spectra are
dominated by the signal-to-noise ratios of just several exposures obtained in
the best seeing conditions.  

\lris\ has a dichroic that splits the spectrum into two channels at $\sim5500\,
{\rm \AA}$. At our resolution and central wavelength, only H$\alpha$, which is
a rather poor mass indicator, landed on the red side.  Hence, the red-side
spectra were not reduced. The blue-side \lris\ spectra were reduced using
standard {\sc iraf} tasks.

\begin{figure*}%~~~~~~~~~~~~~~~~~~~~~~~~~~~~~~~~~~~~~~~~~~~~~~~~~~~~~~~
\includegraphics[width=\textwidth,height=0.99\textheight]{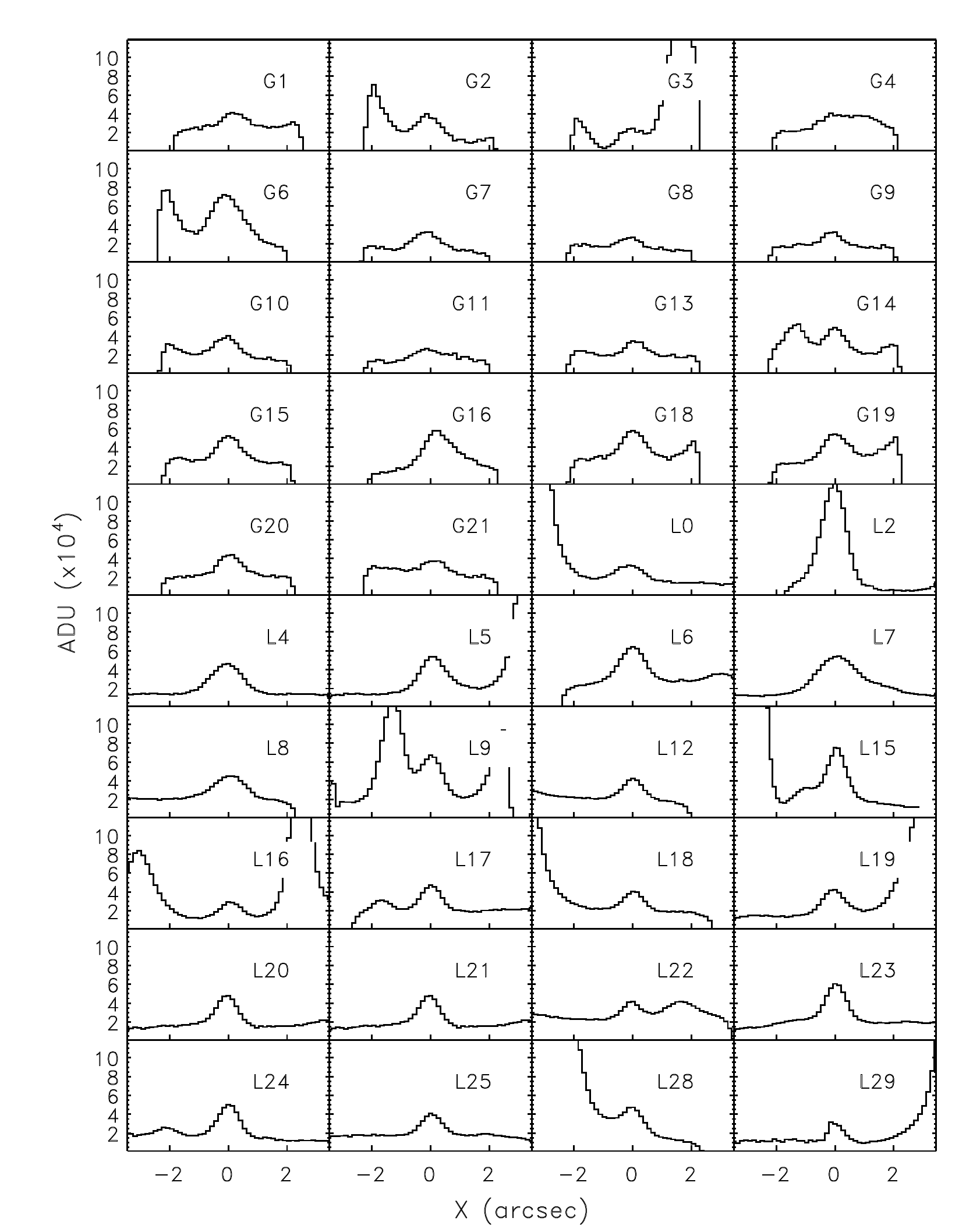} 
\caption{The cross-sections of the slits of the objects obtained with LRIS and
GMOS.  The sections are displayed such that the white dwarf is in the center of
the figure. The cross-sections of the LRIS objects  were calculated from the
first science exposure on July 15, 2007. This was the best exposure of our run
in terms of seeing. The cross-sections of the GMOS objects were calculated from
first science exposure taken of June 6, 2005. \label{ext.fig}}
\end{figure*}

The trace, sky subtraction, and extraction were all performed with the {\sc
apall} task.  \lris\ has better blue sensitivity compared to \gmos, and the
trace at blue wavelengths was therefore far more certain. This field is
extremely crowded. Our ability to obtain a reliable trace, and perform accurate
sky subtraction is dependent on the particulars of each individual slit. The
quality of the trace varies widely from slit to slit. Figure \ref{ext.fig},
shows the flux from the individual slits as a function of spatial pixel,
integrated along the spectral dimension. Note that some of the stars have very
clean slits, and many pixels with which to calculate the sky values (e.g.
LRIS-04), while others have noisy backgrounds or bright stars on the slits
(e.g. LRIS-09). Because of the uneven illumination across the slit due to
nearby neighbours, a simple subtraction of the sky values was insufficient. We
fit a polyline to the sky. The order of the polyline varied between one and
three depending on the shape of the sky. The fit to the sky values was then
interpolated across the pixels containing the signal from the stars, and
subtracted.

The wavelength calibration was calculated from spectra of three lamps
containing Hg, Zn, and Cd. There are not all that many transitions at these
wavelengths, and furthermore, at the resolution we used, some of the lines had
odd shapes. The residuals were not as well distributed as with the \gmos\ data,
and the final dispersion of the wavelength residuals of the line fits was
typically $\sim 0.1\, {\rm \AA}$. While there are not as many transitions as
with the CuAr lamp used for the \gmos\ data, the wavelength solution is
sufficiently precise for our purposes.  The wavelength calibration was
calculated with the {\sc identify} task, and was applied to the science images
with the {\sc dispcor} task.  The flux calibration for these stars was
calculated from the flux standard HZ44, using the default values for the flux.
The response was fix with a spline.  The calibration was calculated using the
tasks {\sc standard}, and {\sc sensfunc}, and were applied to the science
spectra using the task {\sc calibrate}.  Finally, the individual spectra were
combined using {\sc scombine}, with the weights according to their
signal-to-noise ratios.

\begin{figure*}
\includegraphics[width=\textwidth]{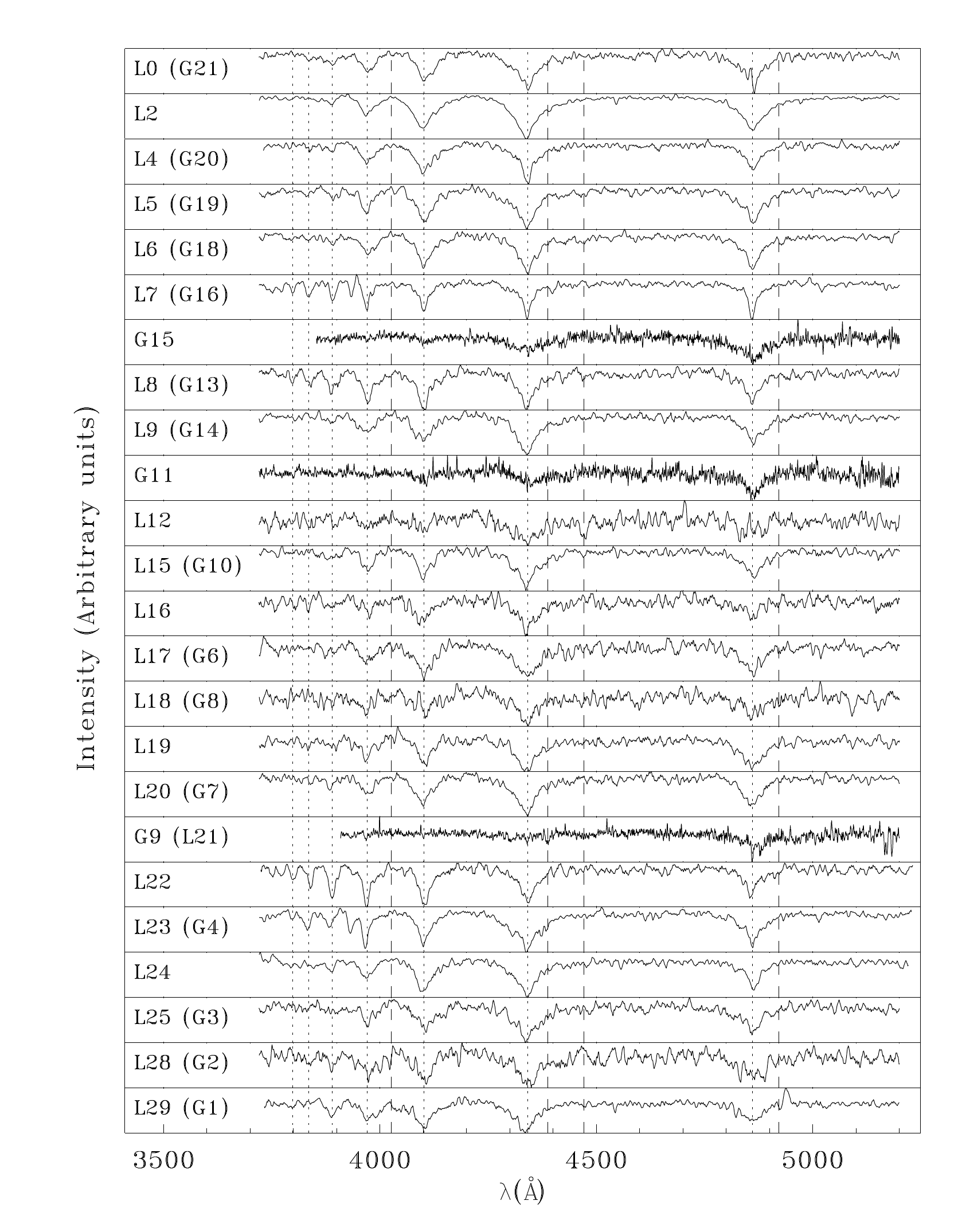}
\caption{\label{spec.fig} The spectra for all confirmed white dwarfs.  Despite
the low signal-to-noise ratio of some of these spectra, they are {\em all}
clearly type DA. The expected position of H${10}$ through H$\beta$ are shown
with dotted lines. The expected position of several helium lines (4026\AA,
4388\AA,
4471\AA, 4922\AA) are shown with
dashed lines. None of the spectra show any significant lines at the positions
of the helium lines. Note, the calcium K-line visible (between H$\varepsilon$
and H$8$) in objects L7 and L23 is most likely contamination due to very
nearby main-sequence stars.}
\end{figure*}

Figure \ref{spec.fig} shows the spectra that could be reliably extracted. While
many of these spectra have too low a signal-to-noise ratio to determine masses,
they all clearly show a Balmer series, and hence can be classified as type DA.
The continuum play no role in the spectral classification, and has therefore
simply been subtracted.  Of the 35 slits observed between the \gmos\ and \lris\
observations, 25 were strong white-dwarfs candidates. The other ten objects
were included to fill the slit mask, but are unlikely to be white dwarfs. These
objects were selected to be CV candidates, or white dwarf--main sequence star
binary candidates.  After preliminary reductions, none of these objects
appeared to be interesting, and hence were not pursued further. All the strong
candidates, except for one object (LRIS-27), have been confirmed to be type DA.
We were unable to extract a reliable spectrum for LRIS-27 due to two other
bright stars on the slit, and hence its spectral type is still unknown. For
postage stamps of the area surrounding each target, see Kalirai et al. 2009.

\section{Discussion}
\subsection{Spectral Types}
In total, we have determined the spectroscopic type of 24 white dwarfs. They
are all of type DA. At this point it is of interest to take a closer look at
the significance of this result.  It is important for the following treatment
that we be able to treat the observed white dwarfs as individual data points.
The \wds s in our sample clearly all have the same metallicity and age,
however, delaying discussion of these points to section \ref{pe.sec}, the
cluster environment is unlikely to impart any other common property.
Correlations of angular momentum, magnetic field, and rotation are all likely
to be washed out during the formation of the cluster \citep{md04}. We can
therefore treat the stars as individual data points.

We can use binomial statistics to determine the probability of observing no DB
white dwarfs. The binomial distribution has the following form:
\begin{equation}
	f(k;n) = { n \choose k}p^k(1-p)^{n-k}%\frac{n!\,p^k(1-p)^{n-k}}{k!(n-k)!}
\label{poisson.eq}
\end{equation}
where $k$ is the number of observed events, $n$ is the number of trials, and
$p$ is the probability of any one event giving a ``positive result'' (i.e.,
having a non-DA spectral type). When no events are observed, Equation
\ref{poisson.eq} takes a particularly simple form:
\begin{equation}
f(0;n) = (1-p)^n
\label{poisson2.eq}
\end{equation}

Under the assumption that the DA/DB ratio, $r$, is the same in the cluster
environment as it is in the field, the probability of any one white dwarf being
of type DB is:
\begin{equation}
p=\frac{1}{1+r}.
\end{equation}

From SDSS-DR4, \citet{elh06} found $r\simeq4.2$ for temperatures between
$15\,000$--$25\,000$ K, implying $p=0.19$.  Putting the preceding equations
together, we have an expression for the probability, $f$, of observing no DB
white dwarfs:
\begin{equation}
f(0;n)=(1-p)^n=0.81^n.
\end{equation}
 
Assuming all 24 white dwarfs are cluster members, the probability of observing
no DB white dwarfs is $6\times10^{-3}$ ($2.5\sigma$ in a normal distribution).
However, there is almost certainly a small level of contamination by field
white dwarfs. Nine of the stars have their proper motions measured from \wfpc\
astrometry. These stars are almost certainly cluster member. The probability of
observing no DBs in this sub-sample is $0.15$ ($1.0\sigma$ in a normal
distribution). An additional ten stars have photometry {\em very} close to the
$0.5\,M_\odot$-cooling curve. If we include these, for a total of 19 cluster
members, the probability of observing no DBs is $0.02$ ($2.1\sigma$ in a normal
distribution). Of the remaining five stars, two (L8 and L22) have photometry
marginally consistent with the white dwarf cooling sequence, while the other
three (L2, L7, and L21) are unlikely to be cluster members. We will only
include nineteen stars with cluster membership established by either proper
motion or photometry in the following analysis.

We should also take into account the white dwarfs identified by \citet{mkz04}.
If we add the 9 white dwarfs identified by them in two other globular
clusters, the total number of globular cluster identifications rises to 28. The
chance of observing no DBs in a sample this large is only $3\times10^{-3}$
($2.8\sigma$ in a normal distribution). As a final note, \citet{sca09} recently
reported the discovery of 24 He-core \wds s in NGC 6397. These all show
H$\alpha$ absorption, and are therefore likely {\em all} DAs. However, because
these are photometric identifications, we will not consider these any further.
It is now clear that the DA/DB ratio in globular clusters is different from
that in the field.

Since 2005, a substantial number of white dwarfs in open clusters have had
their spectral types determined. There are now a handful of non-DA white dwarfs
identified. DBs have been found in NGC 6633 and NGC 6819 by \citet{wb07} and
\citet{khk08} respectively. A DBA has been identified in the Hyades, and a DQ
has been identified in NGC 2168 \citep{wlb06}. It should be noted that none of
the stars have their memberships confirmed by proper motions. However, assuming
that all these stars are indeed genuine cluster members, we can calculate the
probability of observing four or fewer non-DAs, $P(X\le4)$, in the total sample
of $\sim140$ stars: 
\begin{equation}
	P(X\le4)=\sum_{k=0}^4{140 \choose k}(.19)^k(.81)^{140-k}=6\times10^{-9},
	\label{sbinom.eq}
\end{equation}
the equivalent of a $5.7\sigma$ deviation in a normal distribution. It is now
incontrovertible that the DA/DB ratio in the cluster environment is different
from that in the field.

\subsection{Possible Explanations \label{pe.sec}}
From the examination of cooling curves, and from the fitting of spectral lines
(see Kalirai et al. 2009), we can constrain the effective temperatures of our
stars to be cooler than $25\,000$ K. This is cool enough for He-convection to begin
in \wds s with thin hydrogen envelopes, and would therefore transform at least
a fraction of them to DBs. We can therefore use a hydrogen-rich atmosphere as a
proxy for a thick hydrogen layer. The dearth of DB white dwarfs is more
accurately thought of as a dearth of thin hydrogen envelopes. For the remainder
of this section we will refer to \wds s with thick hydrogen envelopes as
hydrogen rich (HR), those with thin hydrogen envelopes as hydrogen poor (HP).

At first blush, one may imagine many explanations that could explain the dearth
of HP \wds s. However, many of these can be eliminated with several simple
observations. The dearth of HP white dwarfs has now been observed in many open
clusters and several globular clusters. These clusters cover a wide range of
metallicity (from $-2.0<$[Fe/H]$<+0.4$), and we will therefore ignore any
explanation that invokes metallicity. Furthermore, these clusters cover an
extremely wide range of ages (from 100 Myr--12 Gyr), and therefore white-dwarf
masses. The dearth of HP \wds s has been seen clearly in populations with
average masses ranging from $M_{WD}\sim0.8$ \msun\ in open clusters
\citep[see][]{krh05} to $M_{WD}\sim0.5$ \msun\ in globular clusters.  These
extreme populations comfortably bracket the peak of the mass distribution
observed in the disk \citep[$\bar{M}_{WD}=0.6$ \msun, see][]{brb95}. We will
therefore also ignore any explanation that invokes white dwarf mass. In the
most general terms, we can propose explanations that either prevent the
formation of HP \wds s, or transform the HP \wds s back to HR ones.

\subsubsection{Suppression of HP formation}

In order to prevent the formation of a HP \wds, we require a mechanism that
will reduce the efficiency with which a star expels or burns its hydrogen
envelope.  One can imagine mechanisms involving binary stars that could affect
the retention of a hydrogen envelope. There is evidence that the binary
fraction in globular clusters is significantly lower than that of the field
\citep[][and references therein]{dra08}. In particular, the binary fraction for
extreme horizontal branch (EHB) stars is much lower in \gc s than it is in the
field \citep{bca08}.  \citet{cc93} showed that if sufficient mass is lost, a
star can depart the red giant branch (RGB) before experiencing a helium core
flash.  These stars will experience the core flash later in their evolution and
are referred to as ``hot flashers'', and will be manifested as EHB stars in a
\gc, or as sub-luminous O stars in the field. \citet{lsh04} showed that if the
flash happens late enough during the star's evolution, the energy of the flash
can drive the expansion of the convection zone so that it engulfs the entire
hydrogen envelope. This hydrogen will then be completely burned, and the
resulting atmosphere will be completely dominated by helium. 

If binaries drive mass loss, and increase the incidence of ``hot flashers'',
then higher binary fraction in the field may account for the increased
incidence of HP \wds s there.  However, only a small fraction of \wds s are
formed from EHB stars, and not all of the EHB stars will be late hot flashers.
This casts doubt upon the binary fraction/EHB channel as begin a likely
explanation. Furthermore, the dearth of HP \wds s is also seen in open
clusters, which have not been shown to have binary fractions significantly
different from the field \citep[see][for example]{kt04, fn08}. This casts doubt
on the binary fraction in general being a feasible explanation.

Another possible avenue to lose mass might be very close encounters with other
cluster members. The rate, $r$, of interactions within a given radius can be
estimated as, 
\begin{equation}
r=n\sigma v,
	\label{nsv.eq}
\end{equation}
where $n$ is the number density of stars, $\sigma$ is the cross-section of
interaction, and $v$ is the typical velocity of the interacting particles.
For the mechanism to be successful, it requires a rate of at least one per
typical RGB-star lifetime, which is assumed to be $t\simeq t_{RGB}=350$ Myr. An
interaction that liberates a significant mass of gas from a RGB star will have
a radius comparable to several times the radius of the star, which is
approximately $R\simeq R_{RGB}=1$ AU. Finally, we assume a velocity that is in
line with a typical velocity dispersion of a star cluster, i.e., $v\simeq
v_5=5\,km/s$.

Rearranging Equation \ref{nsv.eq}, we can estimate the necessary density for
one interaction to occur within the lifetime of a RGB star
\begin{equation}
	n=5\times10^5\,pc^{-3}\,\,
%	t^{-1}_{RGB}\,
%	r^{-2}_{RGB}\,
%	v^{-1}.
\left( \frac{t}{t_{RGB}} \right)^{-1}
\left( \frac{R}{R_{RGB}} \right)^{-2}
\left( \frac{v}{v_5} \right)^{-1}
	\label{n.eq}
\end{equation}
This density is higher than is present in most, if not all, globular clusters,
and {\em far} higher than the majority of open clusters, and is hence unlikely
to be a feasible explanation. 

\subsubsection{Transformation of HP to HR}

Transforming an HP \wds\ to a HR \wds\ simply requires the star to re-accrete a
hydrogen envelope. A re-accreted hydrogen layer can either suppress convection
in the helium layer or, if it is thick enough, resist being drawn into the
interior of the star completely when convection does occur.  It is unclear what
the absolute minimum mass of hydrogen necessary for this, however, we can get a
lower limit by examining the accretion rate of \wds s in the field. The Bondi
accretion rate is
\begin{equation}
	\dot{M}=\frac{4\pi m_p n G^2M^2}{V^3},
	\label{bondi.eq}
\end{equation}
where $M$ is the mass of the star, $V$ is the relative velocity of the gas and
the star, $n$ is the number density of protons, and $m_p$ is the mass of the
proton. Substituting typical values for a disk white dwarf, we have
\begin{equation}
	\dot{M}=5\times10^{-17}\, M_\odot/yr
	\,\frac{M^2_{WD} \,n_{ISM}}{\,V^3_{30}},
	%\,M^2_{WD} \,V^{-3}_{30}\,n_{ISM}.
%	\left(\frac{M}{0.5\,M_\odot}\right)^2
%	\left(\frac{V}{30\,km/s}\right)^{-3}
%	\left(\frac{n}{1\,cm^3}\right)^{}.
	\label{bondiwd.eq}
\end{equation}
where $M_{WD}=0.5\,M_\odot$, $n_{ISM}=1\,cm^{-3}$, and $V_{30}=30\,km/s$. The
stars we are examining have cooling time of approximately $20$ Myr, and hence
we would expect the average field white dwarf at a temperature similar to the
\wds s  observed in M4 to have accreted approximately $10^{-9}\,M_\odot$ of
hydrogen from the ISM. We assume a lower limit to transform a DB to a DA is an
order of magnitude greater than this, i.e., $M_{min} = 10^{-8}\,M_\odot$, or
equivalently $\dot{M}_{min} = 5\times10^{-16}\, M_\odot/yr$, or finally,
$n_{min} = 10\,cm^{-3}$.While we do not yet have a definitive scenario that can
increase the accretion rate to the requisite level, we will briefly examine two
possibilities: accretion from a central reservoir of gas, and accretion from
the cluster wind (i.e., the composite wind from all stars).

An obvious way to increase the density of the intra-cluster medium, and
therefore the accretion rate of the cluster \wds s, is accretion of
gas into the cluster potential. \citet{lm07} explore the accretion of gas
into the gravitational potentials of star clusters, and the subsequent effect
on accretion rates of individual stars. The efficiency at which a cluster can
accrete gas is determined primarily by the relative velocities of the cluster
compared with the ISM. Most globular clusters move at velocities relative to
the ISM of $\sim10^2\, km/s$ with respect to the ISM, and will not effectively
accrete gas. 

As the relative velocity of the cluster and ISM drops below the velocity
dispersion of the cluster, the cluster potential starts to accrete gas, and
dramatically increases the accretion rate of the individual stars. The
estimated rate for a globular cluster with a velocity dispersion of $10\,km/s$,
according to \citet{lm07}, is $10^{-13}\, M_\odot/yr$. This rate is well above
the required rate. However, the fact that most \gc s (including M4, NGC6397,
and NGC6752, i.e.~\ all the globular clusters that have had spectral
observations reported) have velocities well above
this velocity, makes this an unlikely explanation for the observed effect.
Furthermore, the shallow potential of open cluster makes this an unlikely
scenario in that environment too.

Another source of gas could be the collective winds of the other stars in the
cluster. The winds from evolved low- and intermediate-mass stars have typical
velocities of $10$--$30\,km\,s^{-1}$ and hence we would not expect the gas to
be retained by the gravitational potential of most clusters. The accretion
would therefore be from a smooth cluster wind. This scenario was explored for
the open cluster NGC 2099 in \citet{krh05}, and found to be an unlikely
explanation. Following their analysis, we present a more general expression for
the accretion rate due to a cluster wind.

In order to estimate the density of the cluster wind, we make several
assumptions. First, we assume the cluster has a Salpeter mass function from
$0.1$--$8.0\,M_\odot$, the upper limit being the approximate limit at
which \wds s will form.  We determine an expression for the turn-off mass,
$M_{TO}$, as a function of time by examining the models of \citet{dot06}. We
find a satisfactory fit with 
\begin{equation}
	M_{TO}=1.633\left(\frac{t}{1\,{\rm Gyr}}\right)^{-0.2935}.
	\label{mto.eq}
\end{equation}
In order to determine the mass loss per star, we subtract the remnant mass as
determined in Kalirai et al. (2009) from $M_{TO}$. The mass returned to the
ISM, $M_{ISM}$, is
\begin{equation}
	M_{ISM}=0.89\,M_{TO}-0.39\,M_\odot.
	\label{mism.eq}
\end{equation}

The lifetime of a main sequence star combined with the IMF tells us the number
of stars evolving off the main sequence as a function of time. Each star
leaving the main sequence returns a certain mass of gas back to the ISM, and
therefore the total mass begin returned to the ISM as a function of time can be
calculated. Assuming the cluster retains none of this gas, this is equivalent
to the mass loss from the cluster. Unfortunately, due to the form of Equations
\ref{mto.eq} and \ref{mism.eq} this could not be done analytically. The
numerical result was well approximated by the following expression for mass
loss from a cluster due to the composite stellar wind in units of $M_\odot/yr$,
\begin{equation}
	\dot{M}_{cl}=1.22\times10^{-7}\,
	\left(\frac{M_{cl}}{10^4M_\odot}\right) 
	\left(\frac{t+1.1}{1\,Gyr}\right)^{-1.17}.
	\label{ml.eq}
\end{equation}
Of course, not all the stars returning gas to the ISM are at the center of the
cluster, but due to mass segregation the most massive star will be the most
centrally concentrated. The approximation that all the mass loss occurs at the
center of the cluster is good enough for our purposes. The gas density due to
the cluster wind at a radius $R$ from the cluster center will be 
\begin{equation}
	n=1.0\times10^{-3}\, cm^{-3}
	\left(\frac{3\,pc}{R}\right)^2 
	%\left(\frac{\dot{M}_{cl}(10\, Gyr)}{\dot{M}_{cl}(10\, Gyr)}\right) 
	\frac{\dot{M}_{cl}(t)}{\dot{M}_{cl}(1\,Gyr)}.
	\label{dense.eq}
\end{equation}
This density is far below the critical density necessary for accretion to
transform HP to HR \wds s, and is in fact less than the mean density of the
ISM, and hence can be neglected. 

This treatment of mass loss and accretion has been rather crude. We assumed
that the gas leaves the cluster in a smooth way. Perhaps the winds from
different stars interact in a way that causes them to cool, and be retained by
the cluster more efficiently than one would na\"ively expect.

\section{Conclusion}
With an additional 24 white-dwarf spectral-type identifications, we have
roughly tripled the number of spectral identifications in globular clusters.
All the newly identified \wds s are DAs. This extends the already-observed
phenomenon of star clusters being deficient in non-DA \wds s to cover
essentially all metallicities and white-dwarf masses. It is now
incontrovertible that the DA/DB ratio is different in the field than in the
cluster environment. 

The discovery of a handful of non-DA \wds s in several open clusters show that
is not {\em impossible} to form a non-DA white dwarf in the cluster
environment, however, the formation mechanism is clearly strongly suppressed.
Unfortunately, there is no obvious mechanism for this, but it seems likely that
some mechanism exists that enables the re-accretion of a hydrogen envelope. The
dearth of non-DA white dwarfs in clusters relative to the field has now been
clearly demonstrated across the full range of age and metallicity, and remains
an unsolved problem in stellar evolution.

%%fakesection the bibliography
{\it Facilities:} \facility{Gemini:South (GMOS)}, \facility{HST
(WFPC2)}, \facility{Keck:I (LRIS)}
\bibliographystyle{plainnat}

\begin{thebibliography}{26}
\providecommand{\natexlab}[1]{#1}
\providecommand{\url}[1]{\texttt{#1}}
\expandafter\ifx\csname urlstyle\endcsname\relax
  \providecommand{\doi}[1]{doi: #1}\else
  \providecommand{\doi}{doi: \begingroup \urlstyle{rm}\Url}\fi

\bibitem[{Bergeron} et~al.(1997){Bergeron}, {Ruiz}, and {Leggett}]{brl97}
P.~{Bergeron}, M.~T. {Ruiz}, and S.~K. {Leggett}.
\newblock {The Chemical Evolution of Cool White Dwarfs and the Age of the Local
  Galactic Disk}.
\newblock \emph{\apjs}, 108:\penalty0 339, January 1997.
\newblock \doi{10.1086/312955}.

\bibitem[{Bragaglia} et~al.(1995){Bragaglia}, {Renzini}, and {Bergeron}]{brb95}
A.~{Bragaglia}, A.~{Renzini}, and P.~{Bergeron}.
\newblock {Temperatures, gravities, and masses for a sample of bright DA white
  dwarfs and the initial-to-final mass relation}.
\newblock \emph{\apj}, 443:\penalty0 735--752, April 1995.
\newblock \doi{10.1086/175564}.

\bibitem[{Castellani} and {Castellani}(1993)]{cc93}
M.~{Castellani} and V.~{Castellani}.
\newblock {Mass loss in globular cluster red giants - an evolutionary
  investigation}.
\newblock \emph{\apj}, 407:\penalty0 649--656, April 1993.
\newblock \doi{10.1086/172547}.

\bibitem[{Davis} et~al.(2008){Davis}, {Richer}, {Anderson}, {Brewer}, {Hurley},
  {Kalirai}, {Rich}, and {Stetson}]{dra08}
D.~S. {Davis}, H.~B. {Richer}, J.~{Anderson}, J.~{Brewer}, J.~{Hurley}, J.~S.
  {Kalirai}, R.~M. {Rich}, and P.~B. {Stetson}.
\newblock {Deep Advanced Camera for Surveys Imaging in the Globular Cluster NGC
  6397: the Binary Fraction}.
\newblock \emph{\aj}, 135:\penalty0 2155--2162, June 2008.
\newblock \doi{10.1088/0004-6256/135/6/2155}.

\bibitem[{Dotter} et~al.(2006){Dotter}, {Chaboyer}, {Baron}, {Ferguson},
  {Jevremovic}, {Lee}, and {Worthey}]{dot06}
A.~L. {Dotter}, B.~{Chaboyer}, E.~{Baron}, J.~W. {Ferguson}, D.~{Jevremovic},
  H.~{Lee}, and G.~{Worthey}.
\newblock {Self-Consistent Stellar Evolution Models with Updated Physics and
  Variable Abundances}.
\newblock In \emph{Bulletin of the American Astronomical Society}, volume~38 of
  \emph{Bulletin of the American Astronomical Society}, pages 958, December
  2006.

\bibitem[{Eisenstein} et~al.(2006{\natexlab{a}}){Eisenstein}, {Liebert},
  {Harris}, {Kleinman}, {Nitta}, {Silvestri}, {Anderson}, {Barentine},
  {Brewington}, {Brinkmann}, {Harvanek}, {Krzesi{\'n}ski}, {Neilsen}, {Long},
  {Schneider}, and {Snedden}]{elh06}
D.~J. {Eisenstein}, J.~{Liebert}, H.~C. {Harris}, S.~J. {Kleinman}, A.~{Nitta},
  N.~{Silvestri}, S.~A. {Anderson}, J.~C. {Barentine}, H.~J. {Brewington},
  J.~{Brinkmann}, M.~{Harvanek}, J.~{Krzesi{\'n}ski}, E.~H. {Neilsen}, Jr.,
  D.~{Long}, D.~P. {Schneider}, and S.~A. {Snedden}.
\newblock {A Catalog of Spectroscopically Confirmed White Dwarfs from the Sloan
  Digital Sky Survey Data Release 4}.
\newblock \emph{\apjs}, 167:\penalty0 40--58, November 2006{\natexlab{a}}.
\newblock \doi{10.1086/507110}.

\bibitem[{Eisenstein} et~al.(2006{\natexlab{b}}){Eisenstein}, {Liebert},
  {Koester}, {Kleinmann}, {Nitta}, {Smith}, {Barentine}, {Brewington},
  {Brinkmann}, {Harvanek}, {Krzesi{\'n}ski}, {Neilsen}, {Long}, {Schneider},
  and {Snedden}]{elk06}
D.~J. {Eisenstein}, J.~{Liebert}, D.~{Koester}, S.~J. {Kleinmann}, A.~{Nitta},
  P.~S. {Smith}, J.~C. {Barentine}, H.~J. {Brewington}, J.~{Brinkmann},
  M.~{Harvanek}, J.~{Krzesi{\'n}ski}, E.~H. {Neilsen}, Jr., D.~{Long}, D.~P.
  {Schneider}, and S.~A. {Snedden}.
\newblock {Hot DB White Dwarfs from the Sloan Digital Sky Survey}.
\newblock \emph{\aj}, 132:\penalty0 676--691, August 2006{\natexlab{b}}.
\newblock \doi{10.1086/504424}.

\bibitem[{Fontaine} et~al.(2001){Fontaine}, {Brassard}, and {Bergeron}]{fbb01}
G.~{Fontaine}, P.~{Brassard}, and P.~{Bergeron}.
\newblock {The Potential of White Dwarf Cosmochronology}.
\newblock \emph{\pasp}, 113:\penalty0 409--435, April 2001.

\bibitem[{Frinchaboy} and {Nielsen}(2008)]{fn08}
P.~M. {Frinchaboy} and D.~{Nielsen}.
\newblock {The WIYN Open Cluster Study Photometric Binary Survey: Initial
  Findings for NGC 188}.
\newblock In E.~{Vesperini}, M.~{Giersz}, and A.~{Sills}, editors, \emph{IAU
  Symposium}, volume 246 of \emph{IAU Symposium}, pages 109--110, May 2008.
\newblock \doi{10.1017/S174392130801541X}.

\bibitem[{Hansen} and {Liebert}(2003)]{hl03}
B.~M.~S. {Hansen} and J.~{Liebert}.
\newblock {Cool White Dwarfs}.
\newblock \emph{\araa}, 41:\penalty0 465--515, 2003.
\newblock \doi{10.1146/annurev.astro.41.081401.155117}.

\bibitem[{Kalirai} and {Tosi}(2004)]{kt04}
J.~S. {Kalirai} and M.~{Tosi}.
\newblock {Interpreting the colour-magnitude diagrams of open star clusters
  through numerical simulations}.
\newblock \emph{\mnras}, 351:\penalty0 649--662, June 2004.
\newblock \doi{10.1111/j.1365-2966.2004.07813.x}.

\bibitem[{Kalirai} et~al.(2005){Kalirai}, {Richer}, {Hansen}, {Reitzel}, and
  {Rich}]{krh05}
J.~S. {Kalirai}, H.~B. {Richer}, B.~M.~S. {Hansen}, D.~{Reitzel}, and R.~M.
  {Rich}.
\newblock {The Dearth of Massive, Helium-rich White Dwarfs in Young Open Star
  Clusters}.
\newblock \emph{\apjl}, 618:\penalty0 L129--L132, January 2005.
\newblock \doi{10.1086/427551}.

\bibitem[{Kalirai} et~al.(2008){Kalirai}, {Hansen}, {Kelson}, {Reitzel},
  {Rich}, and {Richer}]{khk08}
J.~S. {Kalirai}, B.~M.~S. {Hansen}, D.~D. {Kelson}, D.~B. {Reitzel}, R.~M.
  {Rich}, and H.~B. {Richer}.
\newblock {The Initial-Final Mass Relation: Direct Constraints at the Low-Mass
  End}.
\newblock \emph{\apj}, 676:\penalty0 594--609, March 2008.
\newblock \doi{10.1086/527028}.

\bibitem[{Kalirai} et~al.(2009){Kalirai}, {Davis}, {Richer}, {Bergeron},
  {Catelan}, {Hansen}, and {Rich}]{kdr09}
J.~S. {Kalirai}, D.~S. {Davis}, H.~B. {Richer}, P.~{Bergeron}, M.~{Catelan},
  B.~M.~S. {Hansen}, and R.~M. {Rich}.
\newblock {The Masses of Population II White Dwarfs}.
\newblock \emph{\apj}, 2009.

\bibitem[{Lanz} et~al.(2004){Lanz}, {Brown}, {Sweigart}, {Hubeny}, and
  {Landsman}]{lsh04}
T.~{Lanz}, T.~M. {Brown}, A.~V. {Sweigart}, I.~{Hubeny}, and W.~B. {Landsman}.
\newblock {Flash Mixing on the White Dwarf Cooling Curve: Far Ultraviolet
  Spectroscopic Explorer Observations of Three He-rich sdB Stars}.
\newblock \emph{\apj}, 602:\penalty0 342--355, February 2004.
\newblock \doi{10.1086/380904}.

\bibitem[{Liebert} et~al.(1986){Liebert}, {Wesemael}, {Hansen}, {Fontaine},
  {Shipman}, {Sion}, {Winget}, and {Green}]{lwh86}
J.~{Liebert}, F.~{Wesemael}, C.~J. {Hansen}, G.~{Fontaine}, H.~L. {Shipman},
  E.~M. {Sion}, D.~E. {Winget}, and R.~F. {Green}.
\newblock {Temperatures for hot and pulsating DB white dwarfs obtained with the
  IUE Observatory}.
\newblock \emph{\apj}, 309:\penalty0 241--252, October 1986.
\newblock \doi{10.1086/164595}.

\bibitem[{Lin} and {Murray}(2007)]{lm07}
D.~N.~C. {Lin} and S.~D. {Murray}.
\newblock {Gas Accretion by Globular Clusters and Nucleated Dwarf Galaxies and
  the Formation of the Arches and Quintuplet Clusters}.
\newblock \emph{\apj}, 661:\penalty0 779--786, June 2007.
\newblock \doi{10.1086/515387}.

\bibitem[{M{\'e}nard} and {Duch{\^e}ne}(2004)]{md04}
F.~{M{\'e}nard} and G.~{Duch{\^e}ne}.
\newblock {On the alignment of Classical T Tauri stars with the magnetic field
  in the Taurus-Auriga molecular cloud}.
\newblock \emph{\aap}, 425:\penalty0 973--980, October 2004.
\newblock \doi{10.1051/0004-6361:20041338}.

\bibitem[{Moehler} et~al.(2004){Moehler}, {Koester}, {Zoccali}, {Ferraro},
  {Heber}, {Napiwotzki}, and {Renzini}]{mkz04}
S.~{Moehler}, D.~{Koester}, M.~{Zoccali}, F.~R. {Ferraro}, U.~{Heber},
  R.~{Napiwotzki}, and A.~{Renzini}.
\newblock {Spectral types and masses of white dwarfs in globular clusters}.
\newblock \emph{\aap}, 420:\penalty0 515--525, June 2004.
\newblock \doi{10.1051/0004-6361:20035819}.

\bibitem[{Moni Bidin} et~al.(2008){Moni Bidin}, {Catelan}, and
  {Altmann}]{bca08}
C.~{Moni Bidin}, M.~{Catelan}, and M.~{Altmann}.
\newblock {Is a binary fraction-age relation responsible for the lack of EHB
  binaries in globular clusters?}
\newblock \emph{\aap}, 480:\penalty0 L1--L4, March 2008.
\newblock \doi{10.1051/0004-6361:20078782}.

\bibitem[{Press} et~al.(1986){Press}, {Flannery}, and {Teukolsky}]{pft86}
W.~H. {Press}, B.~P. {Flannery}, and S.~A. {Teukolsky}.
\newblock \emph{{Numerical recipes. The art of scientific computing}}.
\newblock Cambridge: University Press, 1986, 1986.

\bibitem[{Richer} et~al.(2002){Richer}, {Brewer}, {Fahlman}, {Gibson},
  {Hansen}, {Ibata}, {Kalirai}, {Limongi}, {Rich}, {Saviane}, {Shara}, and
  {Stetson}]{rbf02}
H.~B. {Richer}, J.~{Brewer}, G.~G. {Fahlman}, B.~K. {Gibson}, B.~M. {Hansen},
  R.~{Ibata}, J.~S. {Kalirai}, M.~{Limongi}, R.~M. {Rich}, I.~{Saviane}, M.~M.
  {Shara}, and P.~B. {Stetson}.
\newblock {The Lower Main Sequence and Mass Function of the Globular Cluster
  Messier 4}.
\newblock \emph{\apjl}, 574:\penalty0 L151--L154, August 2002.
\newblock \doi{10.1086/342527}.

\bibitem[{Richer} et~al.(2004){Richer}, {Fahlman}, {Brewer}, {Davis},
  {Kalirai}, {Stetson}, {Hansen}, {Rich}, {Ibata}, {Gibson}, and
  {Shara}]{rfb04}
H.~B. {Richer}, G.~G. {Fahlman}, J.~{Brewer}, S.~{Davis}, J.~{Kalirai}, P.~B.
  {Stetson}, B.~M.~S. {Hansen}, R.~M. {Rich}, R.~A. {Ibata}, B.~K. {Gibson},
  and M.~{Shara}.
\newblock {Hubble Space Telescope Observations of the Main Sequence of M4}.
\newblock \emph{\aj}, 127:\penalty0 2771--2792, May 2004.
\newblock \doi{10.1086/383543}.

\bibitem[{Stetson}(1987)]{ste87}
P.~B. {Stetson}.
\newblock {DAOPHOT - A computer program for crowded-field stellar photometry}.
\newblock \emph{\pasp}, 99:\penalty0 191--222, March 1987.

\bibitem[{Strickler} et~al.(2009){Strickler}, {Cool}, {Anderson}, {Cohn},
  {Lugger}, and {Serenelli}]{sca09}
R.~R. {Strickler}, A.~M. {Cool}, J.~{Anderson}, H.~N. {Cohn}, P.~M. {Lugger},
  and A.~M. {Serenelli}.
\newblock {Helium-Core White Dwarfs in the Globular Cluster NGC 6397}.
\newblock \emph{ArXiv e-prints}, April 2009.

\bibitem[{Williams} and {Bolte}(2007)]{wb07}
K.~A. {Williams} and M.~{Bolte}.
\newblock {A Photometric and Spectroscopic Search for White Dwarfs in the Open
  Clusters NGC 6633 and NGC 7063}.
\newblock \emph{\aj}, 133:\penalty0 1490--1504, April 2007.
\newblock \doi{10.1086/511675}.

\bibitem[{Williams} et~al.(2006){Williams}, {Liebert}, {Bolte}, and
  {Hanson}]{wlb06}
K.~A. {Williams}, J.~{Liebert}, M.~{Bolte}, and R.~B. {Hanson}.
\newblock {A Hot DQ White Dwarf in the Open Star Cluster M35}.
\newblock \emph{\apjl}, 643:\penalty0 L127--L130, June 2006.
\newblock \doi{10.1086/505211}.

\end{thebibliography}

\end{document}